**A continuity of Markov blanket interpretations under the Free Energy Principle**

**Anil Seth*, Tomasz Korbak,** and **Alexander Tschantz**
School of Engineering and Informatics
University of Sussex
Brighton, BN1 3DL, UK

Commentary on: Bruineberg, J., Dolega, K., Dewhurst, J., & Baltieri, M. (2022) The Emperor's New Markov Blankets. *Behavioral and Brain Sciences*.

*a.k.seth@sussex.ac.uk

**Abstract.** Bruineberg and colleagues helpfully distinguish between instrumental and ontological interpretations of Markov blankets, exposing the dangers of using the former to make claims about the latter. However, proposing a sharp distinction neglects the value of recognising a continuum spanning from instrumental to ontological. This value extends to the related distinction between 'being' and 'having' a model.

"We should not confuse the foundations of the real world with the intellectual props that serve to evoke that world on the stage of our thoughts". This quote from Ernst Mach ([1], p.531, translated in [2], p.19), surfacing from the origins of the philosophy of science, connects directly to the target article [3], in which Bruineberg and colleagues discuss how Markov blankets (MBs) should be understood within the wider literature of the free energy principle (FEP, [4]), as well as how 'models' and 'modelling' should be interpreted within the cognitive and brain sciences more generally.

MBs are statistical descriptions that partition systems into internal, external, and blanket variables – where the internal variables are conditionally independent of the external variables, given the blanket variables. Bruineberg et al provide a valuable service by distinguishing two interpretations of MBs: 'Pearl blankets' (PBs) and 'Friston blankets' (FBs). PBs embody an instrumental approach, in which MBs are used as tools to aid the analysis of complex systems, for example by identifying sets of variables suitable for further investigation. In contrast, FBs adopt an ontological stance in which they are assumed to either *be* (a literalist reading) – or *usefully approximate* (a realist reading) actually-existing boundaries in the world, such as the boundary between a cell and its milieu, or between an organism or agent and its environment. Bruineberg et al reveal the dangers of conflating these two interpretations, in particular when an instrumental (PB) application is implicitly or explicitly taken to justify ontological (FB) conclusions. Their arguments should be borne in mind by those inclined to help themselves to the FEP to explain their favourite grand mystery, or to take it as gospel.

Having said this, making a sharp distinction is often a useful prelude to recognising a spectrum of positions, each of which may be useful when assessed on its own merits. We suggest this is the case here. For example, one may begin with an instrumentalist approach and progressively refine and extend the corresponding model so as to make increasingly specific claims about the causal mechanisms at play in the system under study – in this way, gradually moving towards a more ontological or realist stance. What does 'refine and extend' mean? It could mean equipping the model with additional features that represent potentially important and context-specific aspects of the relevant boundaries, such as autopoetic (self-producing) processes for biological boundaries [5, 6], and embodied and embedded interactions for cognitive boundaries [7, 8], as well as a recognition of the limited degree to which statistical identification of a MB might generalise to nonequilibrium systems [9, 10].

Bruineberg et al mention these possibilities, but downplay their significance by drawing a contrast between 'additional philosophical assumptions' and 'additional technical assumptions', where the latter implicitly subsumes everything just mentioned. But these modelling strategies and mathematical constraints are more than just additional assumptions, they can often be part-and-parcel of the explanatory model itself. And rather than 'additional philosophical assumptions', what seems to be required is a *recognition* of the model's philosophical context and the claims made on its behalf, so as to avoid the sort of conflation helpfully identified by Bruineberg.

Bruineberg et al worry that, given such additional aspects, whether the MB formalism itself can still be doing any work? The answer is yes, to the extent that it helps specify those parts of a model that are focused on boundaries. By casting the distinction between PBs and FBs as sharp, rather than as extrema on a continuum, Bruineberg et al underestimate the explanatory work that MBs may uniquely be able to do.

Digging a little deeper, one reason we might be tempted to invoke a bright-line distinction between PBs and FBs is because of the dramatic claims made for literalist readings of FBs, in which MBs are seen as really-existing ontological boundaries in physical systems. But – as Mach reminds us – models are always models, whatever their granularity. Once we discount the relevance of an overly literalist reading, the value of a continuity between instrumental and ontological stances becomes easier to appreciate. (Here, it is worth separating Mach's valuable skepticism about literalism from his ultimately doomed project to ground physics solely in phenomenology; it is not likely that Mach would have had much time for FBs, even of a realist flavour.)

The same reasoning can be applied to the distinction between 'being a model' and 'having a model' – a distinction which Bruineberg et al mention, but only in passing (see also [11]). Under the FEP, and following the spirit of the cybernetic pioneers [12], many systems can be interpreted as 'being' a model of their environment. In an example briefly discussed by Bruineberg et al, even a simple Watt governor can be described as performing inference – however it is best thought of not as *having* a model that is used to perform inference, but as *being* a model of its environment, from the perspective of an external observer (see [13] for the original and still instructive version of this argument, in the context of computational theories of mind). By contrast, neurocognitive systems that are modelled as implementing generative models of their sensorium, in order to perform inference through prediction error minimisation, are better described as *having* models, rather than merely *being* models.

This distinction is important, because the status of having (rather than being) a model may speak to a variety of interesting phenomena, such as the potential for counterfactual cognition, imagination and imagery, volitional action of various kinds, and perhaps even the difference between conscious and unconscious perception. Methodologically, the hypothesis that a system *has* a model can be warranted if having that hypothesis leads to novel testable predictions that would not have been made without that hypothesis (see [14] for a related argument). Again, it is beneficial to recognise that this distinction comes in degrees, and that even the (realist, ontological) claim that a system *has* a model should not confuse the map with the territory.

The broader lesson from Bruineberg et al is the need for a healthy interaction across disciplinary boundaries, and especially among philosophy, physics, biology, and cognitive science, in order to avoid the pitfalls of explanatory overreach, and to take advantage of the many opportunities that arise at disciplinary boundaries. Ernst Mach – a physicist who eventually took a Chair in the Department of Philosophy at the University of Vienna, making lasting contributions to psychology and physiology along the way – exemplifies these virtues.

**Acknowledgements**: AKS is supported by a European Research Council Advanced Investigator grant (project CONSCIOUS, grant number 101019254). TK is supported by the Leverhulme Doctoral Scholarship Programme: From Sensation and Perception to Awareness (DS-2017-011). The authors are grateful to the Dr Mortimer and Theresa Sackler Foundation for additional support, and to Chris Buckley and Miguel Aguilera for helpful comments.

**Conflict of interest**: The authors have no conflicts of interest to declare.